\documentclass[]{aa} 
\usepackage{graphicx}  
\usepackage{natbib}
\usepackage{xcolor}
\usepackage{amsmath}   
\usepackage{amssymb}    
\usepackage{makecell}
\usepackage{txfonts}
\usepackage{threeparttable} 
\usepackage{adjustbox}
\usepackage[colorlinks=true,citecolor=blue, linkcolor=blue]{hyperref} 

\newcommand{\wb}[0]{\mbox{WASP-7}\,b}

\newcommand{\cahk}[0]{\ion{Ca}{ii} H and K}
\newcommand{\ha}[0]{\mbox{H$\alpha$}}
\newcommand{\mf}[0]{\texttt{molecfit}}

\newcommand{\hb}[0]{\mbox{H$\beta$}}
\newcommand{\hg}[0]{\mbox{H$\gamma$}}

\bibpunct{(}{)}{;}{a}{}{,}

\begin{document} 

  \title{Transmission spectroscopy of \wb\ with UVES
\thanks{Based on observations collected at the European Southern Observatory under ESO programme ID 091.C-0632(A) (PI: F. Pfeifer).}}

\subtitle{Detection of \ion{Na}{I}~D$_2$ and tentative D$_1$ line absorption}

  \author{Hossein Rahmati\inst{\ref{instBA}} \and
      Stefan Czesla\inst{\ref{instTLS}} \and
      Sara Khalafinejad\inst{\ref{instLSW}}
      \and
      Paul Molli\`{e}re\inst{\ref{instMPS}}}
    
  \institute{Department of physics, Bu-Ali Sina University, Hamedan 65178, Iran\\
\email{h.rahmati@alumni.basu.ac.ir}\label{instBA}
      \and
        Thüringer Landessternwarte Tautenburg, Sternwarte 5, D-07778 Tautenburg, Germany\label{instTLS}
      \and Landessternwarte, Zentrum für Astronomie der Universität Heidelberg, Königstuhl 12, 69117 Heidelberg, Germany\label{instLSW}
      \and Max-Planck-Institut für Astronomie, Königstuhl 17, 69117 Heidelberg, Germany\label{instMPS}} 

\abstract{Transmission spectroscopy is a prime technique to study the chemical composition and structure of exoplanetary atmospheres. Strong excess absorption signals have been detected in the optical \ion{Na}{I} D$1,2$ Fraunhofer lines during transits of hot Jupiters, which are attributed to the planetary atmospheres and allow us to constrain their structure.} 
{We study the atmosphere of \wb\ by means of high-resolution transit spectroscopy in the sodium lines.}
{We analyzed a spectral transit time series of 89 high-resolution spectra of the hot Jupiter WASP-7 b that was observed using the Ultraviolet and Visual Echelle Spectrograph (UVES). We used the telluric lines for an accurate alignment of the spectra and carried out a telluric correction with \mf. Stellar magnetic activity was monitored by investigating chromospheric lines such as the \cahk,\ and hydrogen \ha\ lines. Finally, we obtained transmission spectra and light curves for various lines.}
{The star shows no identifiable flares and, if any, marginal changes in activity during our observing run. The sodium transmission spectra and corresponding light curves clearly show signs of the Rossiter-McLaughlin (RM) effect and the stellar center-to-limb variation (CLV) that we modeled using synthetic spectra. A statistically significant, narrow absorption feature with a line contrast of 0.50\,$\pm$\,0.06\,\% (at $\sim 8.3\sigma$ level) and a full width at half maximum (FWHM) of 0.13\,$\pm$\,0.02\ {\AA} is detected at the location of the \ion{Na}{I}~D$_2$ line. For the \ion{Na}{I}~D$_1$ line signal, we derived a line contrast of 0.13\,$\pm$\,0.04\,\% (at $\sim 3.2\sigma$ level), which we consider a tentative detection.
In addition, we provide upper limits for absorption by the hydrogen Balmer lines (\ha, \hb, and \hg), K~{\sc i} $\lambda$7699\,{\AA}, Ca~{\sc ii} H and K, and infra-red triplet (IRT) lines.}
{}


\maketitle

\section{Introduction}
\label{sect:intro}

\begin{table*}
        \centering
        \begin{threeparttable}
                \label{tbl:orbital parameters}
                \caption{Orbital and physical parameters of the WASP-7 system.}
                \begin{tabular}{llll}
                        \hline\hline
                        \noalign{\smallskip}
                        Parameter & Symbol  & Value & Ref. \\
                        \hline
                        \noalign{\smallskip}
\multicolumn{3}{c}{Stellar parameters} \\
                        \noalign{\smallskip}
        Stellar brightness & V [mag] & 9.51 & 1 \\
                        \noalign{\smallskip}
                        Stellar mass & $M_{*}$ [$M_{\odot}$] & 1.285 $\pm$ 0.063 & 2 \\
                        \noalign{\smallskip}
                        Stellar radius & $R_{*}$ [$R_{\odot}$] & 1.466 $\pm$ 0.094 & 2 \\
                        \noalign{\smallskip}
                        Stellar effective temperature & $T_{\rm eff}$ [K] & 6400 $\pm$ 100 & 2
                        \\
                        \noalign{\smallskip}
                        Projected stellar rotation speed & v\:sin\:i\: [km$s^{-1}$] & 14 $\pm$ 2 & 3
                        \\
                        \noalign{\smallskip}
                        Metallicity & [Fe/H] & 0.0 $\pm$ 0.1 & 2
                        \\
                        \hline
                        \noalign{\smallskip}
\multicolumn{3}{c}{Planet parameters} \\
                        \noalign{\smallskip}
                        Planet Mass &  $M_{\rm p}$ [M$_{\rm Jup}$] & 1.083 $^{+ 0.093} _{-0.088}$ & 2 \\
                        \noalign{\smallskip}
                        Planet Radius & $R_{\rm p}$ [R$_{\rm Jup}$] & 1.363 $\pm$ 0.093 & 2 \\\noalign{\smallskip}
                        Planet surface gravity &  log $g_{\rm P}$ [cgs] & 3.159   $^{+ 0.071} _{-0.068}$ & 2
                        \\
                        \noalign{\smallskip}
                        Equilibrium temperature & $T_{\rm eq}$ [K] & 1530  $\pm$ 45 & 4
                        \\
                        \hline
                        \noalign{\smallskip}
\multicolumn{3}{c}{Transit parameters} \\
                        \noalign{\smallskip}
                        Orbital Period & $P$ [d] &  4.9546416 $\pm$ 3.5 $\times$ 10$^{-6}$ & 3
                        \\
                        \noalign{\smallskip}
                        Mid-transit & $T_{0}$ [BJD$_{TDB}$ $-2,400,000$] &  55446.635 \  $\pm$ 0.0003 & 3 \\\noalign{\smallskip}
                        Transit duration & $T_{14}$ [h] & 4.12 $^{+ 0.09} _{- 0.06}$ & 3 \\
                        \noalign{\smallskip}
                        Ingress duration & $T_{12}$ [min] & 27 $^{+ 6} _{- 9}$ & 3 \\
                        \hline
                        \noalign{\smallskip}
\multicolumn{3}{c}{System parameters} \\
                        \noalign{\smallskip}
Orbital inclination & $i$ [deg] &  87.03  $\pm$ 0.93 & 5
                        \\
                        \noalign{\smallskip}
                        Semimajor axis & a\: [AU] & 0.06188  $^{+ 0.00098} _{-0.001}$ & 2 \\
                        \noalign{\smallskip}
                        Stellar velocity semi-amplitude & $K_{\star}$ [m\,s$^{-1}$] & 109.4 $^{+ 8.3} _{-8.1}$ & 2
                        \\
                        \noalign{\smallskip}
                        Systemic velocity & $\gamma$ [km\,s$^{-1}$] & -29.35 $\pm$ 0.53 & 6
                        \\
                        \noalign{\smallskip}
                        Projected obliquity & $\lambda$ [deg] & 86 $\pm$ 6 & 3
                        \\
                        \noalign{\smallskip}
                        \hline
                \end{tabular}
                \tablebib{
                (1)~\citet{Hellier2009}; 
                (2) \citet{Bonomo2017};  
                (3) \citet{Albrecht2012}; 
            (4) \citet{Southworth2012};
            (5) \citet{Southworth2011};
            (6) \citet{Gaia}.
}
        \end{threeparttable}
\end{table*}

To date, more than 5000 exoplanets have been discovered with a wide variety of masses and orbital configurations. The observed diversity raises questions regarding planetary formation, evolution, and the state of the planetary atmospheres. The latter, in particular, holds the promise to evaluate their potential to host life. Many of the known exoplanets transit their host stars. This special orbit configuration enables the use of transmission spectroscopy to study their atmospheres, which has proved to be a highly successful technique. During the transit, a fraction of the stellar light passes through the atmosphere, which leaves its footprint on it in the form of absorption by atomic and molecular species. The study of this light is the subject of transmission spectroscopy.

Hot Jupiters, which are gaseous giant planets orbiting their host stars at close proximity, are promising targets for transmission spectroscopy as their atmospheres can cover a particularly large fraction of the disks of their host star. Hot and ultra-hot Jupiters have equilibrium temperatures of $\sim$ 1500 k and higher than 2000 K, respectively \citep{Glidic2022}. The equilibrium temperature of a planet can be obtained using Eq. (\ref{eq:Teqplanet}) by \citet{Mendez2017}
:

\begin{equation} \label{eq:Teqplanet}
T_{eq} = T_{\star} \sqrt{\frac{R_{\star}}{2d}} \left( \frac{1-A}{\beta \epsilon} \right) ^\frac{1}{4} \; ,
\end{equation}

where $T_{\star}$ is the effective temperature of the star, $d$ is the distance of the planet from the star, $A$ is the planetary albedo, and $\epsilon$ is the broadband thermal emissivity ($\epsilon \approx 1$). The parameter $\beta$ is the fraction of the planetary surface that re-radiates the absorbed flux, which is equal to one and 0.5 for quick rotators and tidally locked planets without an atmosphere, respectively \citep{Kaltenegger2011}.

In fact, pronounced absorption signatures of \ion{Na}{i} and other alkali metals were predicted to show in optical transmission spectra by early models \citep[][]{SeagerSasselov2000, Brown2001} and, indeed, the first detection of \ion{Na}{i} in-transit excess absorption in the atmosphere of HD 209458\,b soon followed \citep{Charbonneau2002}, though it has not remained unchallenged \citep[][]{CasasayasBarris2020, CasasayasBarris2021}. The first reports of ground-based detections of in-transit excess absorption in the \ion{Na}{i}~D lines were provided by \citet{Redfield2008} and \citet{Snellen2008}, who studied the atmospheres of HD~189733\,b and HD~209458\,b, respectively. High-resolution transmission spectroscopy has allowed the individual \ion{Na}{i} transmission line profiles to be resolved \citep[e.g.,][]{Wyttenbach2015, Wyttenbach2017, Seidel2019}, which makes the \ion{Na}{i} signatures a particularly powerful probe to constrain the upper atmosphere \citep[][]{Wyttenbach2015, Nikolov2018}.

The hot Jupiter, \wb,\ was discovered using the transit method by \citet{Hellier2009}. The planet orbits an F-type star, \object{HD 197286,} with an orbital period of 4.954~d. The main properties of the system are summarized in Table~\ref{tbl:orbital parameters}. Notably, the orbit of \wb\ denotes an obliquity of $86^{\circ} \pm 6^{\circ}$, indicating an almost polar orbit \citep{Albrecht2012}. This special configuration makes the WASP-7 system a promising candidate to study differential rotation via the planetary transit \citep{Serrano2020}.

In this paper, we present our analysis of a high-resolution spectral transit time series of \wb, obtained with the Ultraviolet and Visual Echelle Spectrograph (UVES) at the Very Large Telescope (VLT). The structure of this paper is as follows. In Sect.~\ref{sec:obs}, we present the observations and data reduction. In Sect.~\ref{sect:dataanalysis}, we investigate the stellar activity, focus on \ion{Na}{I}~D lines, and provide upper limits of plausible absorption for other lines. Finally, we present a discussion and conclusion in Sects.~\ref{sect:discussion} and \ref{sect:conclusion}.

\section{Observations and data reduction} 
\label{sec:obs}

A high-resolution spectral time series of \wb\ was obtained on 31 August 2013 in the context of ESO programme 091.C-0632(A) using UVES \citep[][]{Dekker2000}, mounted on the Kueyen telescope (UT2), which is part of the VLT. We downloaded the data from the ESO archive and carried out the data reduction using standard recipes from the UVES pipeline. In particular, the pipeline performs bias subtraction, flat fielding, and wavelength calibration. Finally, the spectra are merged into 1D science spectra. With our settings, the UVES instrument covers the wavelength range from 380 to 950~nm.

\begin{figure}[ht]
  \includegraphics[width = \columnwidth, height= 0.6\columnwidth]{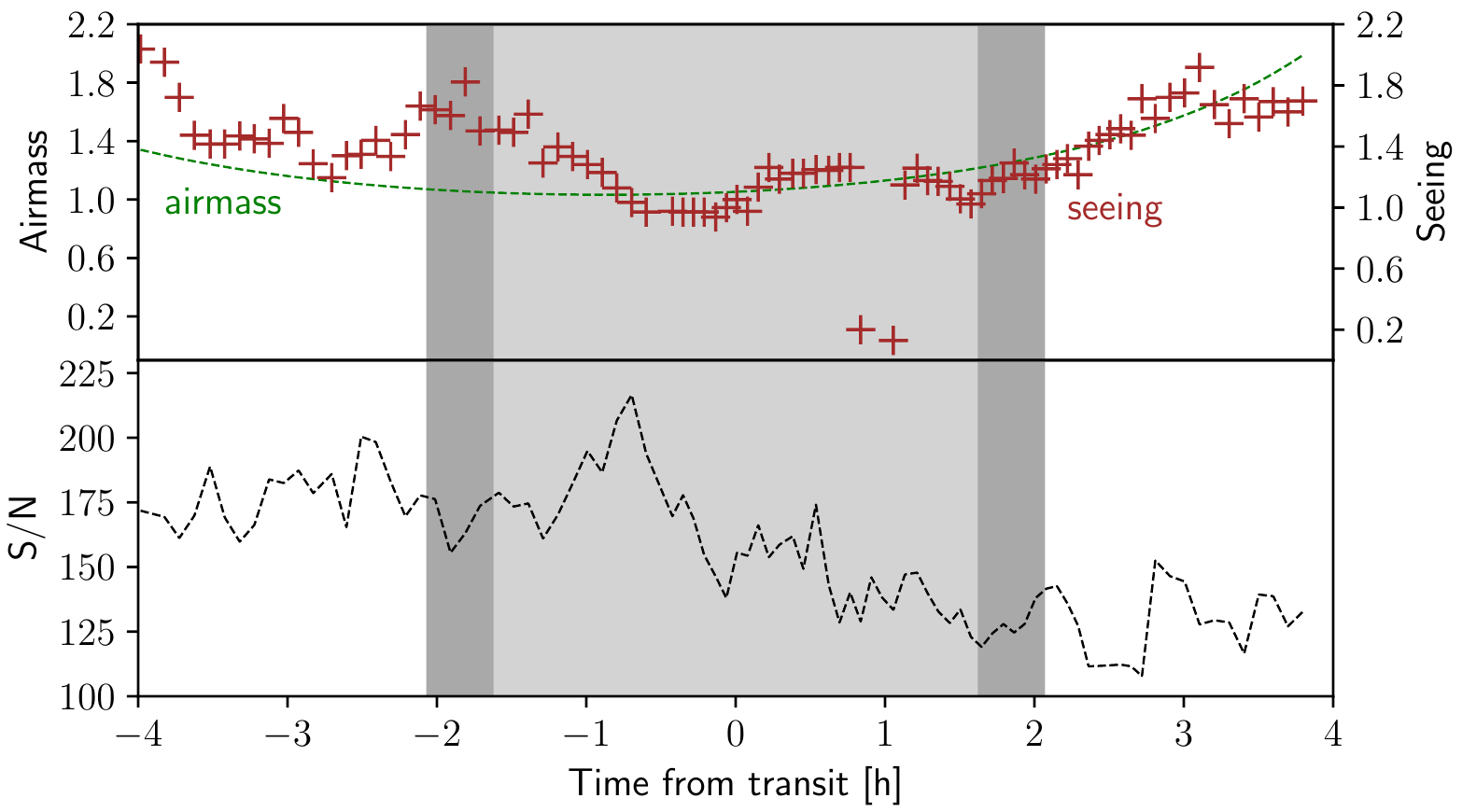} 
  \caption{Time evolution of seeing in arcsec (brown crosses, top panel), airmass (dashed green lines, top panel), and S/N in the regions of the Na doublet lines (bottom panel). Dark gray shades indicate the ranges of ingress and egress, and light shades show the range between second and third contact.
  \label{fig:airmass}}
\end{figure}

The time series comprises 89 spectra with a spectral resolution of $60\,000$ in the \ion{Na}{i} regions. A total of 20 exposures were obtained before ingress; 49 during the transit and 20 after egress. The first 34 exposures, exposure 48, and exposures 79-89 have an exposure time of 300~s, while the remaining exposures have an exposure time of 200~s. In the top panel of Fig.~\ref{fig:airmass}, we show the evolution of airmass and seeing during the observing run. The airmass ranges between 1.03 and 1.98 during the observing run. The average seeing is 1.24~arcsec and its evolution roughly follows that of airmass. The bottom panel of Fig.~\ref{fig:airmass} displays the time evolution of the signal-to-noise ratio (S/N) per spectral bin in the region of the Na lines, estimated from the data using the $\beta\sigma$ procedure \citep[][]{Czesla2018}. The average S/N value of the spectra is about 150, starting higher at about 178 when the run starts, after which it slightly decreases toward the end of the observing run.

\subsection{Telluric correction}

\begin{figure}[ht]
  \includegraphics[width=\columnwidth,height= 0.6\columnwidth]{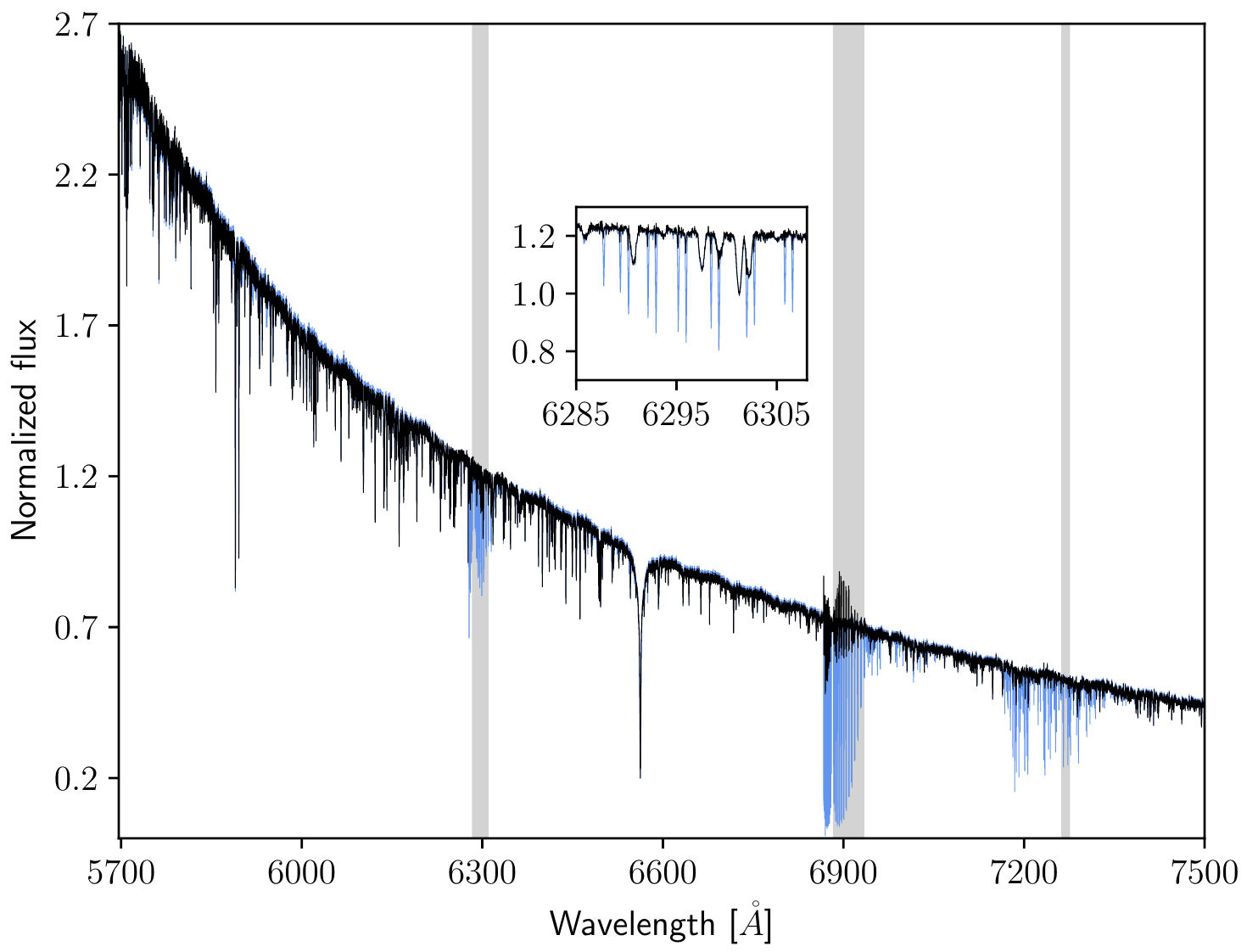} 
  \caption{Spectrum before (blue) and after (black) telluric correction.
  The gray shaded regions denote sections used for fitting with \mf.}
  \label{fig:telluricCorrection}
\end{figure}

Ground-based high-resolution spectra are affected by the features originating in the Earth’s atmosphere, and changes in airmass and atmospheric conditions along the line of sight cause variations in the telluric lines. We corrected the telluric lines in the \ion{Na}{I}~D, \ha,\ K~{\sc i}, and Ca~{\sc ii} infra-red triplet (IRT) line regions using the \mf\ package \citep[][]{Smette2015, Kausch2015}. The \mf\ package fits a synthetic transmission spectrum of the Earth’s atmosphere to suitable sections of the observed spectrum and, subsequently, corrects the telluric contamination in the observed spectrum using the synthetic model evaluated across the entire observed range. Clearly, the regions of the \ion{Na}{I} and \ha\ lines are mostly affected by water lines. In the \cahk\ line region, telluric lines remain irrelevant. As an example of our telluric correction, we show the last spectrum of our time series in the \ion{Na}{I}~D and \ha\ regions in Fig.~\ref{fig:telluricCorrection}. This exposure exhibits the largest airmass and is most strongly affected by telluric absorption (Fig.~\ref{fig:airmass}, top panel).

\subsection{Spectral alignment}
\label{sec_alignment}

\begin{figure}[ht]
  \includegraphics[width = \columnwidth, height= 0.6 \columnwidth]{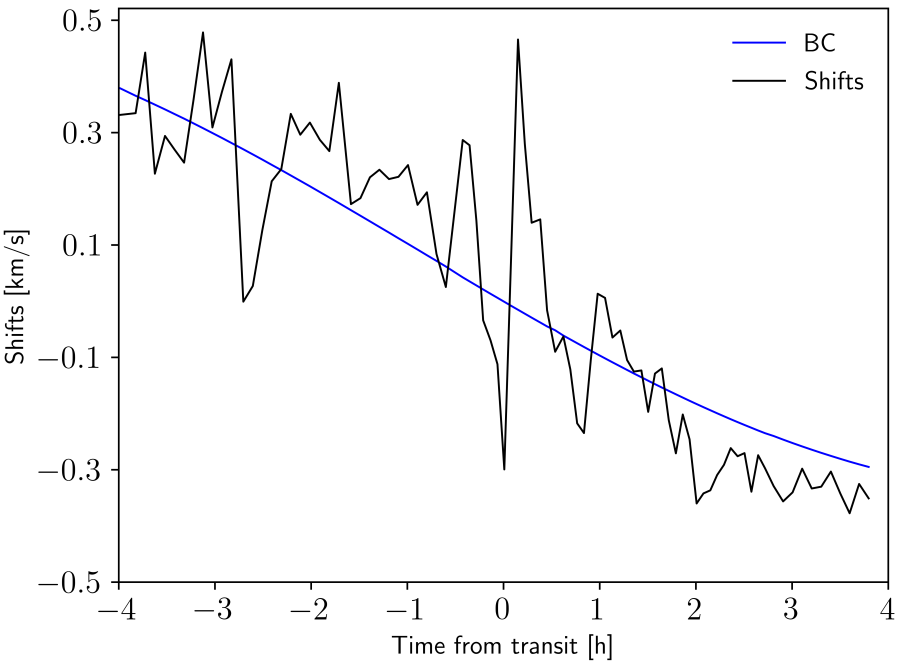} 
  \caption{Time evolution of BC and median telluric shifts used for spectral alignment.}
  \label{fig:shifts}
\end{figure}

In high-resolution transmission spectroscopy, an accurate alignment of the spectra is crucial because misalignment can produce remnants in the transmission spectrum. The dispersion in UVES is known to depend on pressure and temperature \citep[e.g.,][]{Czesla2012}. Therefore, we used the wavelengths of 39 telluric lines, which provide a reference for the Earth's system of rest, accurate to within about $25$~m\,s$^{-1}$ \citep[e.g.,][]{Caccin1985, Gray2006}, to refine the alignment. In Fig.~\ref{fig:shifts}, we show the median shift of the telluric lines along with the evolution of the barycentric correction (BC). We also accounted for the stellar orbit velocity, which remains small, taking values between $-22$ and $22$~m\,s$^{-1}$ during the observing run. Finally, we considered a systemic shift of $-29.35$~km\,s$^{-1}$ for the WASP-7 system \citep[][]{Gaia}. To test the accuracy of our alignment, we fit Gaussian profiles to the cores of the \ion{Na}{I} lines after the alignment. The results are consistent with static line cores and a radial velocity (RV) standard deviation of $10$~m\,s$^{-1}$.

\section{Data analysis}
\label{sect:dataanalysis}
\subsection{Stellar activity}

Stellar activity is a common nuisance in transmission spectroscopy. Stellar flares can produce strong spectral evolution in the cores of chromospherically sensitive lines such as the \cahk\ lines, the corresponding infrared triplet, the \ha\ line, and notably also the \ion{Na}{I}~D lines on a timescale of minutes to hours \citep[e.g.,][]{Klocova2017}, posing serious problems to planetary transmission spectroscopy \citep[e.g.,][]{Barnes2016, Khalafinejad2017}. In addition, stellar surface features such as starspots and plage regions can influence the transmission spectrum \citep[e.g.,][]{Oshagh2014, Cauley2018, Salz2018}. 

\begin{figure} [ht]
  \includegraphics[width = \columnwidth, height= 0.85 \columnwidth]{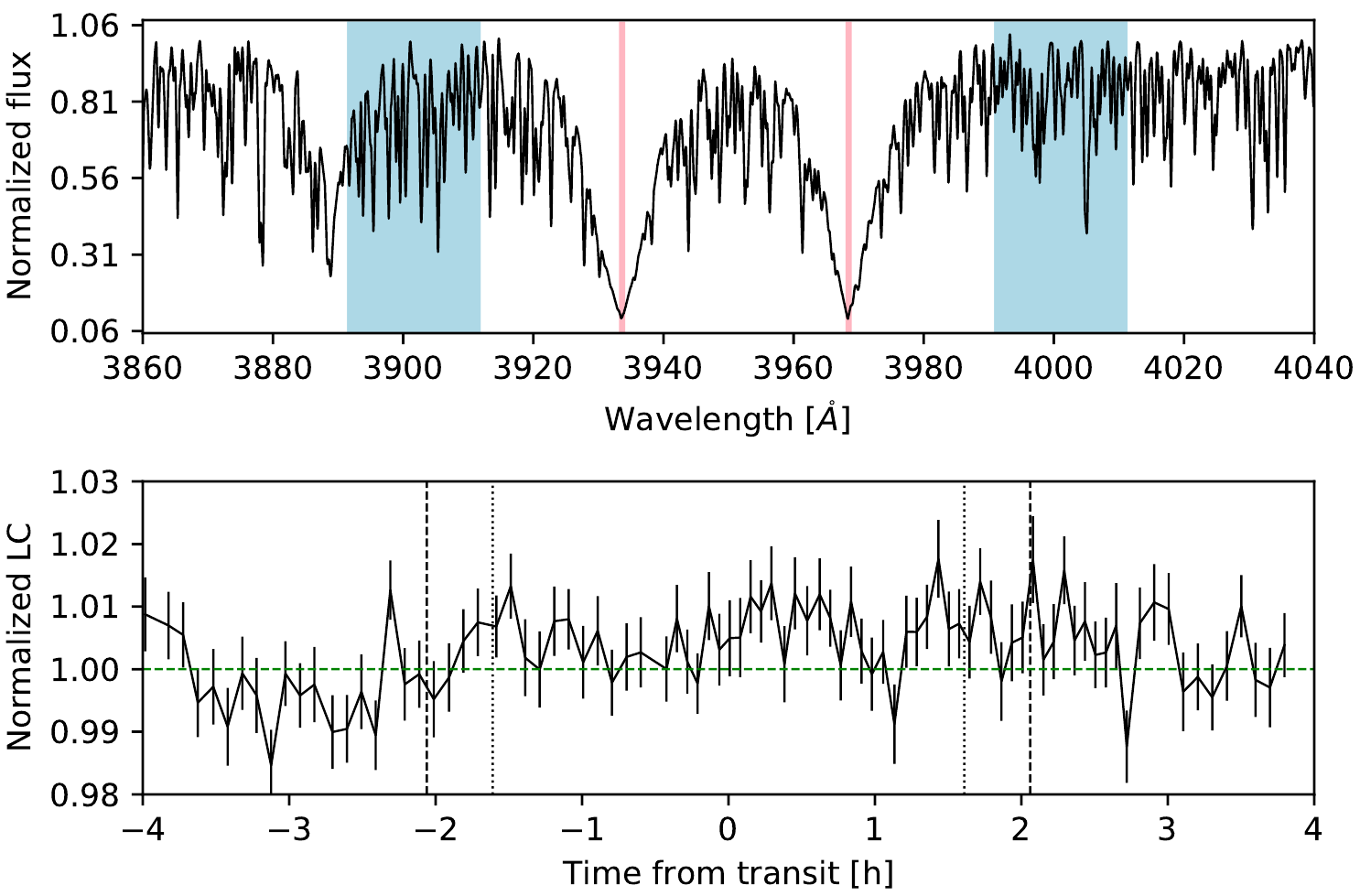} 
  \caption{Integration and reference bands in the \cahk\ line region (top), and normalized \cahk\ light curves. Dashed vertical lines show first and fourth contact, and the dotted vertical lines show second and third contact.}
    \label{fig:Ca}
\end{figure}

\begin{figure} [ht]
  \includegraphics[width = \columnwidth, height= 0.85 \columnwidth]{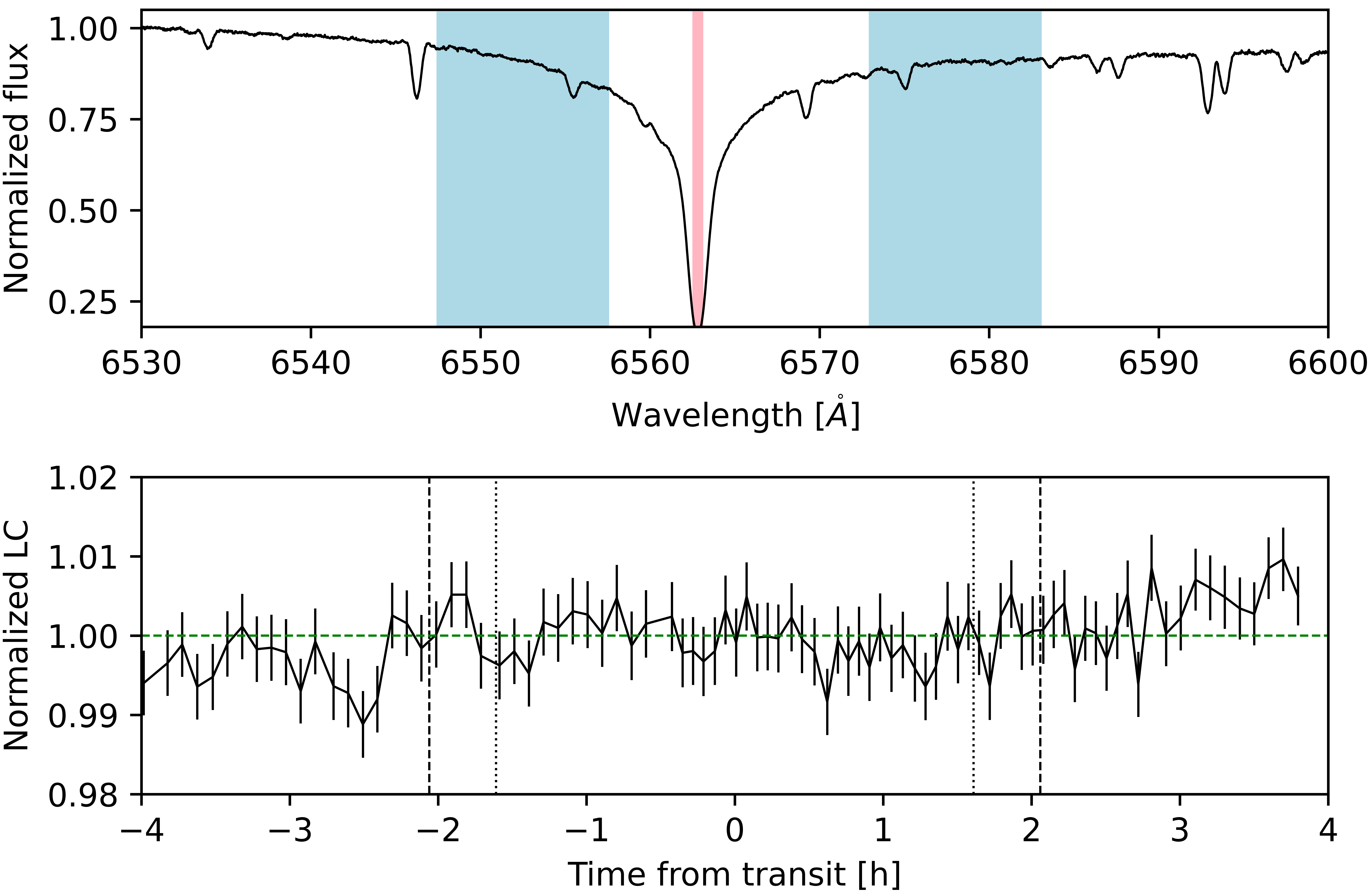}
  \caption{Integration and reference bands in the \ha\ line region (top) and the normalized \ha\ light curve. Dashed vertical lines indicate first and fourth contact, and the dotted vertical lines indicate second and third contact.}
\label{fig:H_alpha}
\end{figure}

We investigated the \cahk,\ and H$\alpha$ lines to search for signs of evolution in activity, such as flaring. In particular, we integrated the flux in a $0.45$~\AA\ wide passband centered on each of the \cahk\ lines, and in broader reference bands on both sides (see Fig.~\ref{fig:Ca}, top panel), and we derived a light curve by dividing the fluxes in the integration and reference bands. A comparable analysis was carried out for the \ha\ line (see Fig.~\ref{fig:H_alpha}). The resulting light curves, shown in the bottom panels of Figs.~\ref{fig:Ca} and \ref{fig:H_alpha}, reveal no discernible flaring activity and are consistent with a constant level of activity with, at most, marginal evolution.

\begin{figure}[ht]
  \centering
  \includegraphics[width = \columnwidth, height= 0.6 \columnwidth]{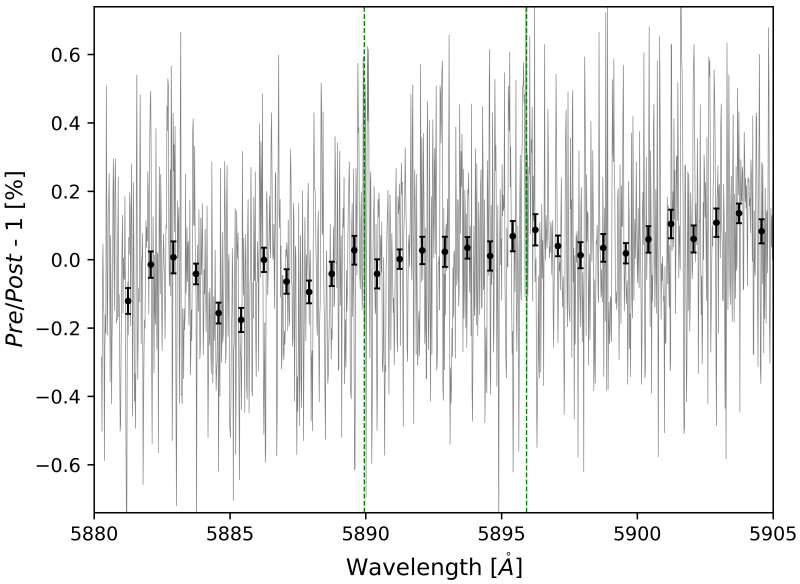}
  \caption{Ratio of pre- and post-transit master spectra (light gray) along with binned spectral ratio (black circles).
  Dashed vertical green lines indicate the centers of \ion{Na}{I} lines.}
  \label{fig:prepostmaster}
\end{figure}

\subsection{Master and residual spectra}

In a first step, we obtained separate weighted pre- and post-transit master spectra by averaging the respective spectra of the time series. The ratio of these spectra around the \ion{Na}{I} lines is displayed in Fig.~\ref{fig:prepostmaster}. No significant relics can be distinguished at the wavelengths of the \ion{Na}{I} line cores, which might indicate potential shortcomings of our alignment (Sect.~\ref{sec_alignment}) or traces of activity. We then proceeded to obtain a weighted master out-of-transit spectrum (OOT master) by averaging all out-of-transit spectra and dividing all spectra of the time series by this OOT master to produce the residual spectra.

The residual spectra can be used to study spectral changes with respect to the OOT master. Such changes may be brought about by absorption in the planetary atmosphere, as well as variability in the stellar lines and statistical effects. The planetary atmospheric signal, in particular, is expected to move through the stellar line along with RV caused by the planetary orbital motion. Therefore, to study the planetary signal, we produced residual spectra by shifting all in-transit residual spectra into the planetary frame:

\begin{equation}
R =  \frac{F_{\mathrm{in}}(\lambda)} { F'_{out}}(\lambda) \rvert _{\mathrm {\:\textit{Planet\: RV\: shift}}} - 1 \; ,
\label{eq-residuals}
\end{equation}

where

\begin{equation}
F'_{out}(\lambda) = \frac{\sum_{} F_{out} W_{\mathrm{S/N}}({F_{\mathrm{out}}(\lambda)})} {\sum_{} W_{\mathrm{S/N}}({F_{\mathrm{out}}(\lambda)})} \; ,
\label{eq-residuals2}
\end{equation}

and ${F'_{out}(\lambda)}$ is the weighted OOT master, ${F_{out}(\lambda)}$ denotes the out-of-transit spectra, and $W_{\mathrm{S/N}}({F_{\mathrm{out}}(\lambda)})$ represents the weights, which are based on the S/N of the individual out-of-transit exposures.

\subsection{Center-to-limb variation and Rossiter-McLaughlin effects}
\label{sec:CLV}

The center-to-limb variation (CLV) describes the spectral variation of the stellar disk as a function of the limb angle. During the transit, the opaque planetary disk subsequently covers different sections of the rotating stellar surface. The combined action of rotational shift and CLV gives rise to pseudo-signals in the transmission spectrum, unrelated to the planetary atmosphere \citep[e.g.,][]{Khalafinejad2017,Salz2018,CasasayasBarris2020}. Such pseudo-signals can be particularly pronounced in strong absorption lines such as the \cahk\ or \ion{Na}{i} lines \citep[e.g.,][]{Czesla2015, Yan2015}.

\begin{figure}[ht]
  \centering
  \includegraphics[width = \columnwidth, height= 0.6 \columnwidth]{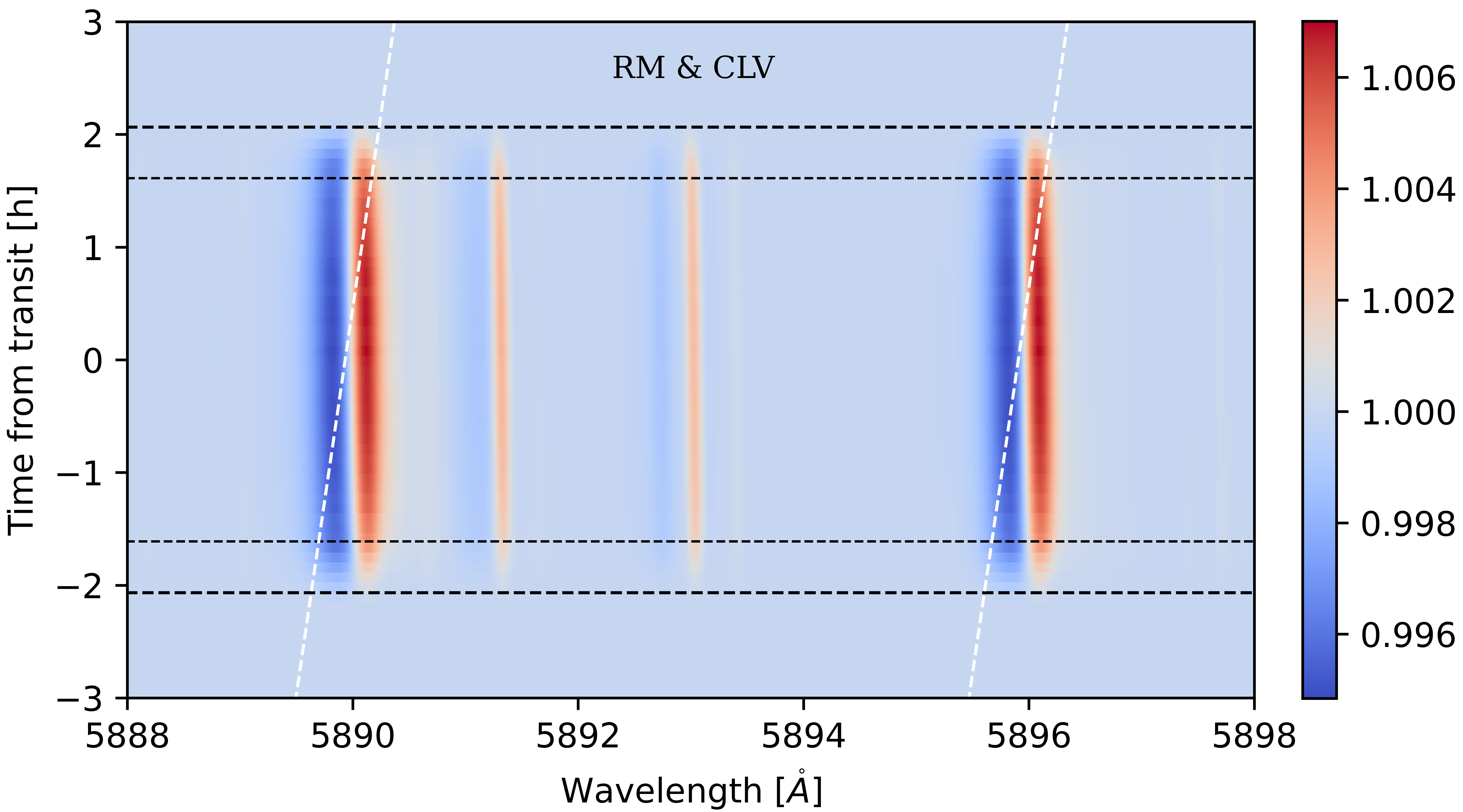}
  \caption{2D heat map of the modeled CLV and RM effects of WASP-7 b around the \ion{Na}{i} D doublet denoted in the stellar rest frame. Horizontal dashed and dotted lines indicate the transit contact, and the dashed white lines show the planetary RV track.}
  \label{fig:2D-simulation}
\end{figure}

To account for the pseudo-effects in our analysis, we simulated the time series of transmission spectra resulting from the transit of the opaque planetary disk alone, following the methodology presented by \citet{Czesla2015}. In particular, we used a discretized, rotating stellar surface \citep{Vogt1987} and limb-angle resolved model spectra of the stellar disk, synthesized using the \texttt{spectrum} \footnote{\url{https://www.appstate.edu/~grayro/spectrum/spectrum.html}}
program by R.O.~Gray \citep[e.g.,][]{Gray1994} based on Kurucz atmospheric models \citep[e.g.,][]{Castelli2003}. Blocking the light from the respective surface elements covered by the advancing planetary disk yields the theoretical spectrum at any point in time during the transit, which is then compared with the disk-integrated stellar spectrum to assess the expected pseudo-signals. In Fig.~\ref{fig:2D-simulation}, we indicate the modeled spectral time evolution caused by the CLV and the Rossiter-McLaughlin (RM) effects as a 2D heat map; the resulting light curve and transmission spectrum are shown in Figs.~\ref{fig:transmission-light-curve} and \ref{fig:transmission-spectrum}, respectively.

\subsection{Transmission light curves}
\label{sec:tlc}

To derive transmission light curves for the \ion{Na}{I}~D lines, we integrated the shifted residual spectra in 0.45~{\AA} wide passbands, centered on the nominal positions of the \ion{Na}{I}~D lines, and compared that result to the flux in suitable reference passbands within 15~{\AA} of the \ion{Na}{I}~D lines' center. The light curve was then normalized by the mean out-of-transit level.

\begin{figure}[ht]
        \includegraphics[width=\columnwidth]{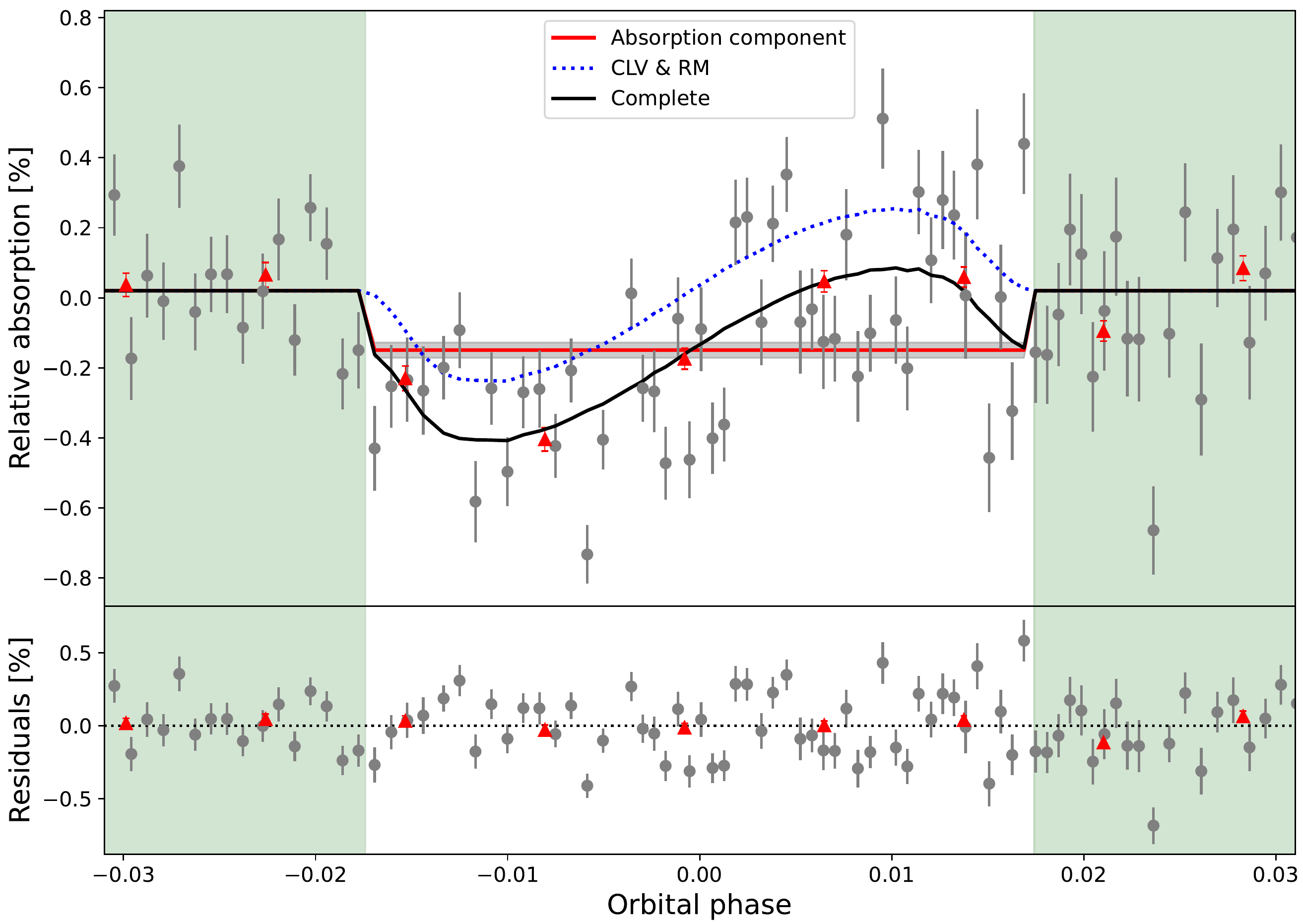}
        \caption{Transmission light curves of WASP-7 b around the \ion{Na}{i}~D$_2$ line. Top panel: Observation transmission light curves (gray) and modeled transmission light curves of the RM and the CLV (dotted blue line) for the 0.45~\AA\ integration band centered on the \ion{Na}{I}~D$_2$ line. The dashed red line is the absorption component using a box model, and the gray shades denote its 1$\sigma$ uncertainty. The solid black line is the best-fit combined model and the red points show binning by a factor of ten. Bottom panel: Residuals for binned and unbinned data. Green shades indicate out-of-transit time throughout.}
        \label{fig:transmission-light-curve}
\end{figure}

In Fig.~\ref{fig:transmission-light-curve}, we show the light curves acquired in this way for the stronger \ion{Na}{I}~D$_2$ line.
We modeled the light curves using increasingly complex models, and show the best-fit results and the values of the Bayesian Information Criterion \citep[BIC,][]{Schwarz1978} in Table~\ref{tab:lcbic}.
Uncertainties were obtained by Markov chain Monte Carlo (MCMC) sampling \footnote{We used the \texttt{emcee} \citep{ForemanMackey2013} and \texttt{PyAstronomy} packages \citep{pya}.}. We startd by using a constant, $c$, and subsequently added a slope, $s$. According to the BIC, the introduction of a slope is not merited, so we dropped that component. Adding the simulated pseudo-signals caused by the CLV and RM effects to the model improved the BIC by $109.1$, providing very strong evidence for the component \citep{KassRaftery1995}. Finally, we included a box-like transit model, which represented in-transit absorption in the \ion{Na}{I}~D$_2$ line potentially attributable to the planetary atmosphere. The best-fit depth, $b$, of the absorption box model was $(1.7 \pm 0.25)\times 10^{-3}$ and its introduction improved the BIC by $39.1$ compared to the previous model, which again provides strong evidence. In Fig.~\ref{fig:transmission-light-curve}{\bf ,} we show the final model, along with its individual components in the top panel and the resulting residuals in the bottom panel.

As an additional test, we estimated the uncertainty of the depth of the box-like transit model using the jackknife method, the idea of which is to analyze the response of the fit to leaving out individual data points \citep[e.g.,][]{Efron1981}. The resulting estimate for the depth was
$(1.7 \pm 0.5)\times 10^{-3}$. The larger error estimate resulting from applying the jackknife method may indicate that the errors of the data were underestimated. Adopting the more conservative jackknife estimate, which takes into account the empirical distribution of the data, we obtained $3.4\,\sigma$ for the significance of the absorption component.

\begin{table}
  \centering
    \caption{Best-fit parameters of the \ion{Na}{I}~D$_2$ line transmission light curve,
    along with $68$\,\% credibility intervals and BIC.
    \label{tab:lcbic}}
    \begin{tabular}{l l}
    \hline \hline
    \multicolumn{2}{c}{Constant} \\
    $c$ & $0.99910\pm 0.00012$ \\
    BIC & 466.6 \\
    \hline
    \multicolumn{2}{c}{Constant and slope} \\
    $c$ & $0.99908\pm 0.00012$ \\
    $s$ [$\AA^{-1}$] & $-0.0049\pm 0.007$ \\
    BIC & 470.6 \\ 
    \hline
    \multicolumn{2}{c}{Constant and CLV} \\
    $c$ & $0.99919\pm 0.00012$ \\
    BIC & 357.5 \\ 
    \hline
    \multicolumn{2}{c}{Constant, CLV, and box} \\
    $c$ & $1.00020\pm 0.00020$ \\
    $b$ & $0.00170\pm 0.00025$ \\
    BIC & 318.4 \\ 
    \hline
    \end{tabular}
\end{table}

\subsection{The transmission spectrum}
\label{sect:TSspectra}

We obtained the transmission spectrum as the weighted average of the shifted residual spectra between the second and third contact:

\begin{equation}
\Re' \:(\lambda) = \frac{\sum_{} R_{\mathrm{in}} W_{\mathrm{S/N}}({F_{\mathrm{in}}(\lambda)})} {\sum_{} W_{\mathrm{S/N}}({F_{\mathrm{in}}(\lambda)})} \; ,
\label{eq-TS2}
\end{equation}

where $R_{\mathrm{in}}$ are the in-transit residuals and $W_{\mathrm{S/N}}({F_{\mathrm{in}}(\lambda)})$ represents the weights, which are based on the S/N of the individual exposures.

\begin{figure}[ht]
        \centering
  \includegraphics[width = \columnwidth, height= 0.6 \columnwidth]{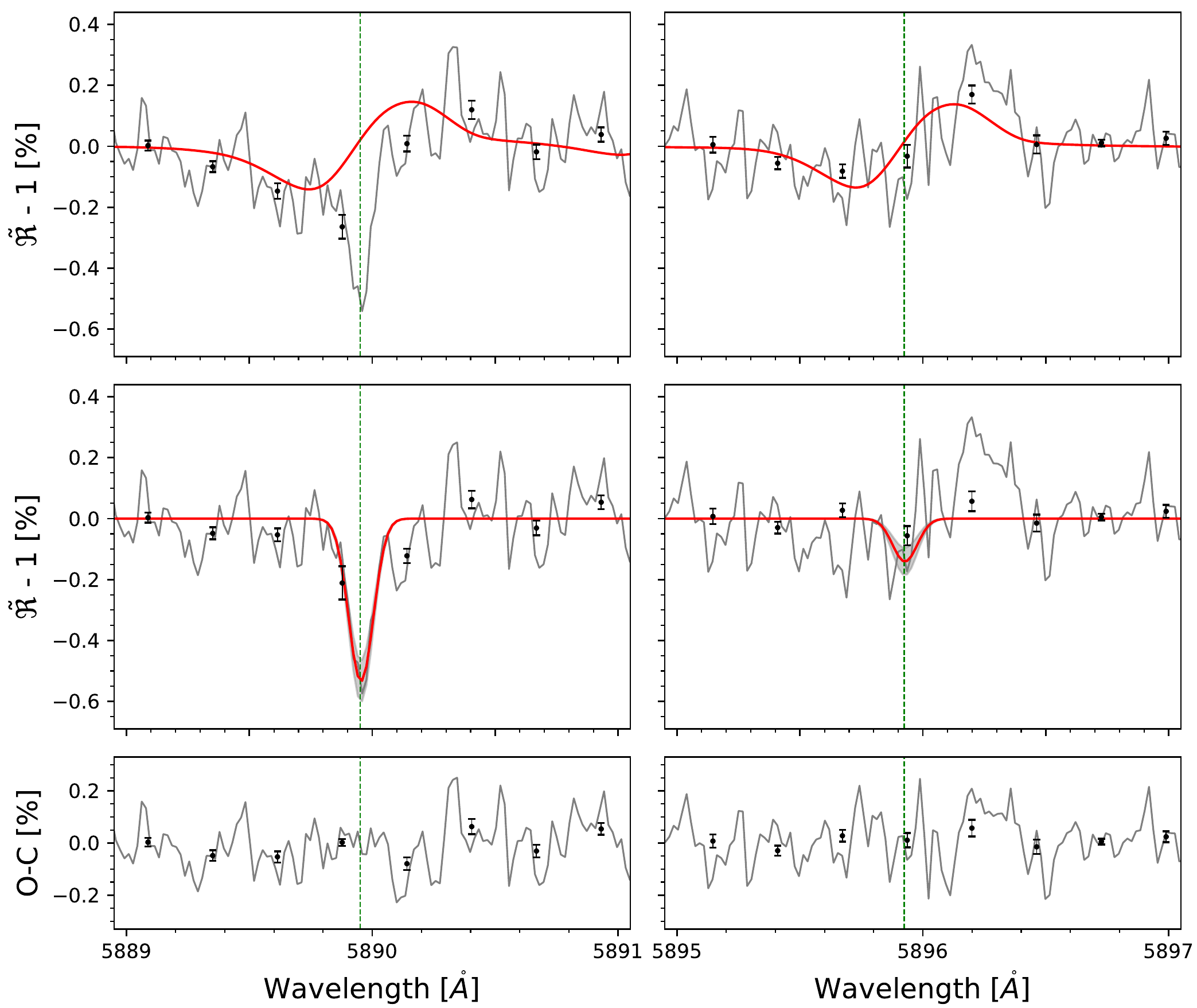}
        \caption{Transmission spectrum of WASP-7 b around the \ion{Na}{i}~D$_2$ (left column) and \ion{Na}{i}~D$_1$  (right column) lines. Top panels: Observed transmission spectrum (gray), along with best-fit model components representing the CLV and RM effects (solid red). Middle panels: Observed transmission spectrum with the best-fit Gaussian absorption components (red), and with the model shown in the upper panels subtracted (gray). The gray shades indicate 1$\sigma$ uncertainty of the best-fit model (red). Bottom panel: Residuals with respect to the best-fit model. The black data points correspond to binning by a factor of 15. 
        }
        \label{fig:transmission-spectrum}
\end{figure}

As demonstrated, for instance, by \citet{Czesla2015} and \citet{CasasayasBarris2020}, the  effect of the CLV and the RM effects on the transmission spectrum can be significant. Therefore, we derived a synthetic transmission spectrum (Sect.~\ref{sec:CLV}), which exclusively shows the effect of the CLV and the RM effects and is indicated in the upper panel of Fig.~\ref{fig:transmission-spectrum}, along with the data.

\begin{figure}[ht]
  \centering
  \includegraphics[width = \columnwidth, height= 0.6 \columnwidth]{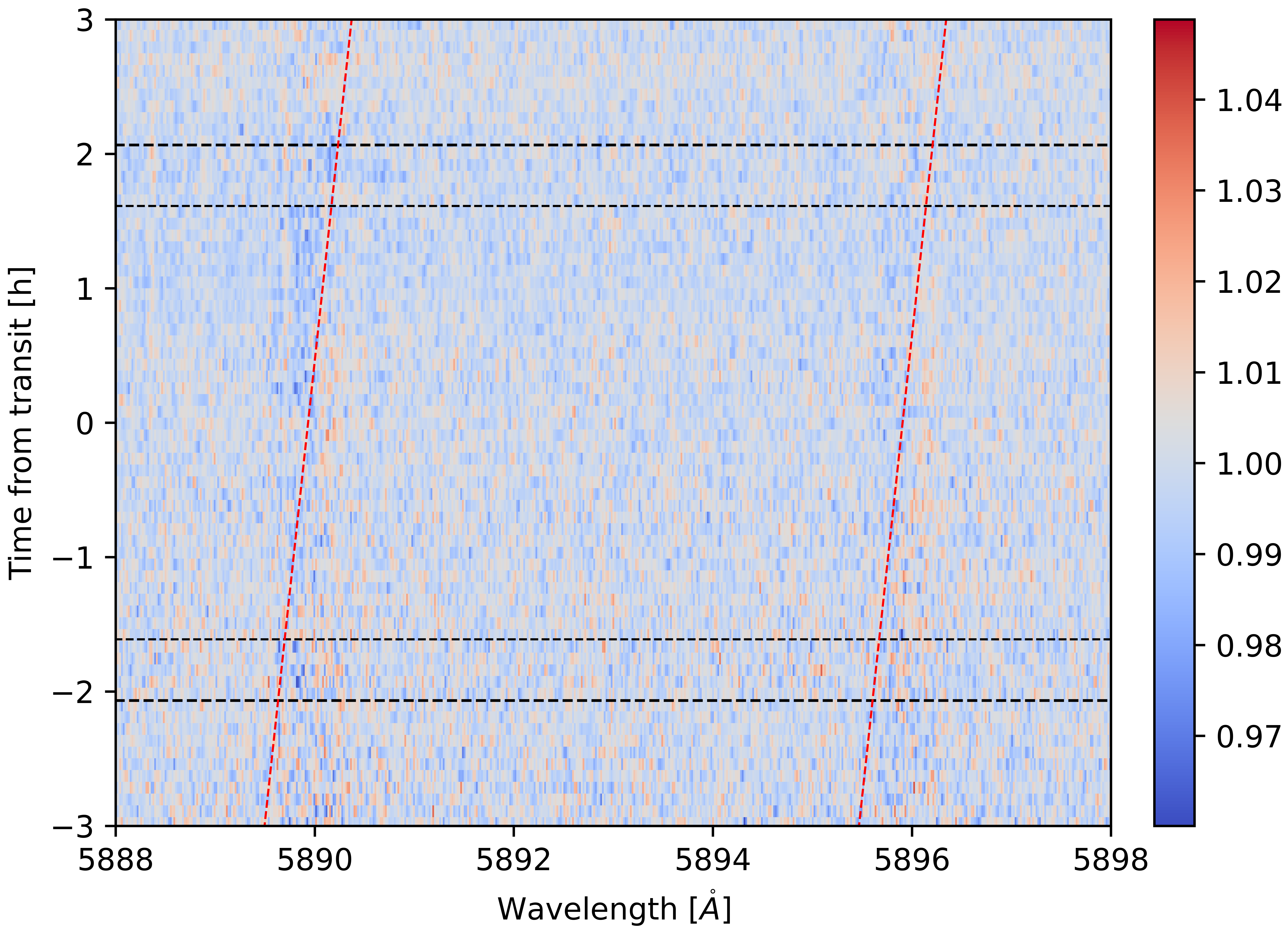}
  \caption{2D heat map of the transmission spectra of WASP-7 b around the \ion{Na}{i} D doublet indicate in the stellar rest frame without considering the CLV and RM effects. Horizontal dashed and dotted lines show the transit contact and the dashed red lines denote the planetary RV track.}
  \label{fig:2D-residuals}
\end{figure}

We modeled the observed transmission spectrum in the range from $5880$~\AA\ to $5905$~\AA\ using the sum of three components: (1) a linear model with a free offset ($p_0$) and slope ($p_1$) to account for possible slopes in the transmission spectrum; (2) the synthetic transmission spectrum with a free scaling parameter ($s$) to account for systematic differences between the true stellar spectrum and the adopted synthetic spectrum of the stellar photosphere; and (3) two Gaussian components, representing possible planetary absorption components in the \ion{Na}{i}~D$_1$ and D$_2$ lines. While the two Gaussian components were set up to show the same width and RV shift with respect to the nominal rest wavelengths of the \ion{Na}{i}~D$_1$ and D$_2$ lines, we treated their strengths as independent. The free parameters of the Gaussian model were, therefore, a common RV shift and standard deviation, and the area of the Gaussians. We then applied Powell's method to find the maximum likelihood solution and, subsequently, carried out MCMC sampling, using uniform priors for the parameters, to find credibility intervals\footnote{We used \texttt{SciPy} \citep{scipy}, \texttt{PyAstronomy} \citep{pya}, and \texttt{emcee} \citep{ForemanMackey2013}.}. The best-fit parameters, along with their uncertainties, are given in Table \ref{tab:tsbic}. In Fig.~\ref{fig:2D-residuals}, we indicate the time evolution of the spectral residuals in the form of a 2D heat map, and in Fig.~\ref{fig:transmission-spectrum}, we show the observed transmission spectrum along with our best-fit model.

\begin{table}
  \centering
        \caption {Best-fit parameters along with $68$\,\% credibility intervals of
        the transmission spectrum model. Values of the BIC for four variants of the model (see text).}
         \label{tab:tsbic}
    \begin{tabular}{l l l}
    \hline \hline
    Parameter & Value & Unit \\ \hline
    $p_0$ & $0.0138\pm 0.0035$ \\
    $p_1$ &  $0.0016\pm 0.0003$ & [$\AA^{-1}$] \\
    $s$ & $0.435\pm 0.040$ \\
    RV & $0.32\pm 0.41$ &  [km$s^{-1}$] \\
    Contrast D$_2$ & $0.500 \pm 0.064$ &  [\%] \\
    Contrast D$_1$ & $0.134 \pm 0.043$ &  [\%] \\ 
    FWHM & $0.135 \pm 0.025$ & [$\AA$]  \\
    \hline
    Model & k & BIC \\ \hline
    No s & 6 & 1554.5 \\ 
    No Abs & 3 & 1571.0 \\ 
    Only~D$_2$ & 6 & 1437.9 \\ 
    Only~D$_1$ & 6 & 1580.3\\
    D$_{1}$ and D$_{2}$ & 7 & 1434.1  \\
    \hline
    \end{tabular}
\end{table}

The transmission spectrum in Fig.~\ref{fig:transmission-spectrum} indicates an absorption feature at the position of the \ion{Na}{I}~D$_2$ line. According to our modeling, the line contrast (i.e., the depth) of the \ion{Na}{I}~D$_2$ component is 0.5\,$\pm$\,0.06\,\%, which corresponds to an $\sim 8.3\sigma$ detection. For the  \ion{Na}{I}~D$_1$ line, we find a best-fit contrast of 0.13\,$\pm$\,0.04\,\% and, thus, a signal at the $3.2\sigma$ level. The best-fit value of the full width at half maximum (FWHM) is 0.13\,$\pm$\,0.02\ {\AA}, which is slightly larger than the instrumental resolution, indicating a narrow intrinsic absorption component. We find a RV shift of $ 0.3\pm 0.4$~km\,s$^{-1}$ for the Gaussian components, consistent with a signal at the rest wavelength. The line shift and width are clearly driven by the D$_2$ line signal.

To further assess the significance of the absorption line components, we calculated the BIC for a total of four model configurations (Table~\ref{tab:tsbic}). First, no scaling pseudo-effects was examined (No s). Second, no sodium absorption was investigated at all (No Abs). Then, absorption was only considered in the D$_2$ and D$_1$ lines individually (Only D$_{1,2}$). Finally, the BIC was calculated for the full model with absorption in both lines (D$_{1}$ and D$_{2}$). In 
Table~\ref{tab:tsbic}, we also give the respective number of free model parameters ($k$). As the widths and contrasts of the Gaussians were coupled, only one additional parameter was introduced by going from a model with a single line to both absorption lines. In line with the error analysis, the BIC provides strong evidence for D$_2$ absorption, but not D$_1$ absorption. 

In an attempt to improve the observed transmission spectrum, we constructed an alternative transmission spectrum by ignoring the ten spectra with the lowest S/N in the observing run, which corresponded to the post-transit exposures 70-80. We then performed the same modeling and obtained a line contrast of 0.49\,$\pm$\,0.05\,\% ($\sim 9.8\sigma$) and 0.19\,$\pm$\,0.03\,\% ($\sim 6.3\sigma$) for the \ion{Na}{I}~D$_2$ and D$_1$ lines, respectively, with a FWHM of 0.17\,$\pm$\,0.02\ {\AA}. This yielded a more significant absorption feature in the \ion{Na}{I}~D$_1$ line.

The detection of the \ion{Na}{i}~D$_2$ feature is statistically significant at $\sim 8.3\sigma$ level between the second and third contact. As a cross-check, we also obtained the transmission spectra corresponding to the first and second halves of the transit individually. The \ion{Na}{i}~D$_2$ line contrasts in the first and second halves of the transit, acquired from a fit with line width fixed to the value given in Table~\ref{tab:tsbic}, are 0.53\,$\pm$\,0.04\,\% and 0.49\,$\pm$\,0.08\,\%, respectively. The feature is slightly more pronounced in the first-half transmission spectrum of the transit, which may be related to the better S/N during the first half. We derived an average contrast of 0.51\,$\pm$\,0.06\,\%, which is consistent with the result of the \ion{Na}{i}~D$_2$ line contrast in Table~\ref{tab:tsbic}.

\subsection{Relation to the transmission light curve}

The equivalent width (EW) of the Gaussian fit to the \ion{Na}{i}~D$_2$ signal is $0.69\pm 0.14$~m\AA\ (Table~\ref{tab:tsbic}). Therefore, the mean depth (in percent) across a band of width $w$ (given in \AA), fully comprising the signal, can be calculated as $(0.069\pm 0.014)\, w^{-1}$. Adopting $w = 0.45$~\AA\ as in Sect.~\ref{sec:tlc}, we find an average depth of $0.15\pm 0.03$\,\%, which is consistent with the result of the light curve analysis.

In light of the spectral results, we also constructed a transmission light curve in a narrower $0.15$~\AA\ wide band. From the same analysis as in Sect.~\ref{sec:tlc}, we find a depth of $(0.309\pm 0.084) \times 10^{-3}$ for the respective box model, which is in line with a narrow signal. While no scaling of the simulated pseudo-effects was indicated in the light curve analysis (Sect.~\ref{sec:tlc}), scaling does improve the fit to the transmission spectra (Table~\ref{tab:tsbic}). We speculate that the different behavior is related to the much wider wavelength range of the transmission spectrum sensitivity considered in this analysis and the resulting change in sensitivity to the model.

\subsection{Upper limits for absorption lines}
\label{sect:upper-limit}

We carried out the same analysis to derive the transmission spectrum of the hydrogen Balmer lines (\ha, \hb, and \hg), K~{\sc i} $\lambda$7699\,{\AA}, \cahk, and the IRT lines. No absorption feature associated with these lines was detected. We computed the standard deviation of the continuum in the regions of the lines and considered that as an upper limit for the line contrast (Table~\ref{tab:upper-limit}).

\begin{table}[ht]
  \centering
    \caption{Upper limits of lines in the transmission spectrum of \wb\ with UVES.}
    \label{tab:upper-limit}
    \begin{tabular}{l l}
    \hline \hline
    Line & Value \\ 
    \hline
   \ha & 0.13 \%  \\
   \hb & 0.12 \%  \\
   \hg & 0.12 \%  \\
   K~{\sc i} $\lambda$7699\,{\AA} & 0.10 \%  \\
   \cahk & 0.16 \%  \\
   Ca~{\sc ii} IRT & 0.11 \%  \\
    \hline
    \end{tabular}
\end{table}

\section{Discussion}
\label{sect:discussion}

\subsection{Comparison with the atmospheric model}
\label{sec.modeling}

To model the transmission spectrum of \wb, we used the \texttt{petitRADTRANS} (pRT) python package \citep[][]{mollierewardenier2019}, which enables the
calculation of transmission and emission spectra of exoplanets at low and high resolution. We adopted the line-by-line mode for the sodium lines and considered Rayleigh scattering by H$\sb{2}$. Collision-induced absorption was ignored because its opacity does not have a critical effect at the observed wavelength. We further adopted a solar sodium abundance, a mean molecular weight of $2.3$, and an isothermal atmosphere with the equilibrium temperature given in Table \ref{tab.parameters} to set up the atmospheric model. The pressure range of the atmosphere between 100 and $10^{-12}$ bar was sampled with 100 layers. We checked that the setup was sufficient by calculating the transmission contribution function in the center of the line core, which showed that the contribution of lower pressure layers was negligible. Finally, we created a model transmission spectrum using the radius, temperature, and surface gravity of the planet at a reference pressure, $P_0$, of 0.1~bar.

\begin{figure}[ht]
        \centering
  \includegraphics[width = \columnwidth, height= 0.6 \columnwidth]{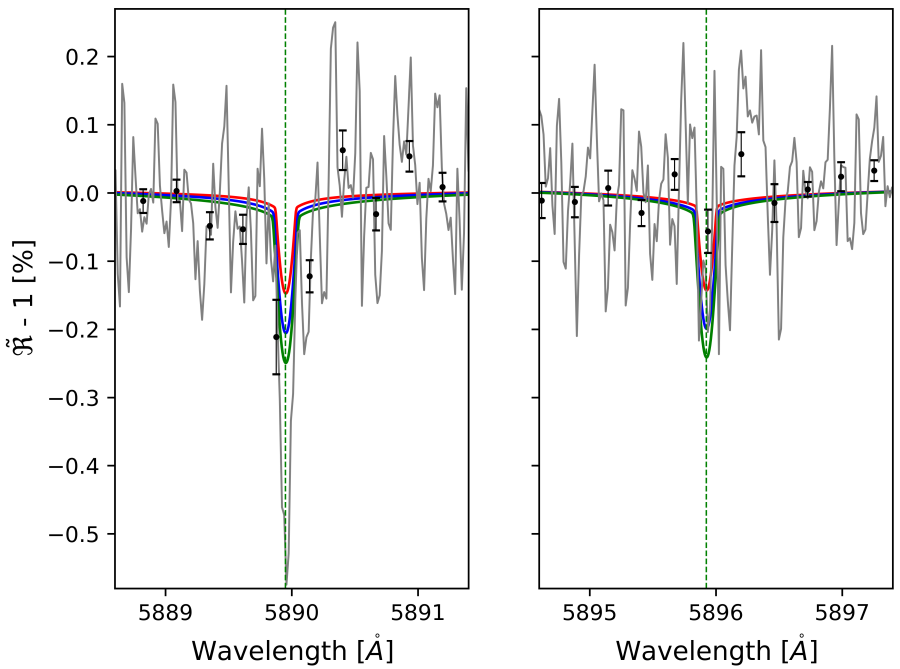}
        \caption{Transmission spectrum of WASP-7 b (gray) around the \ion{Na}{i}~D$_2$ (left column)
        and \ion{Na}{i}~D$_1$ line (right column), along with atmospheric models (red, blue, and green).}
        \label{fig:W7models}
\end{figure}

The resulting model, accounting also for instrument resolution, is shown as the red profile in Fig.~\ref{fig:W7models}. The model produces absorption in the \ion{Na}{I}~D$_2$ line. The line wings seen in the model are undetectable in our data. Approximating the modeled line by a Gaussian, we find a line contrast of 0.15\,\% and a FWHM of 0.14\ {\AA}. The model reproduced the observed width of the line core, but the line is weaker. The modeled \ion{Na}{I}~D$_2$ line strength can, for example, be increased by assuming a higher temperature or a lower mean molecular weight, effectively increasing the atmospheric scale height. Therefore, we also constructed alternative models. We increased the temperature and sodium abundance, first by 30 \% (blue line in Fig.~\ref{fig:W7models}), and then by 50 \% (green profile). The observed \ion{Na}{I}~D$_2$ line contrast was still not reproduced. We speculate that the assumption of hydrostatic equilibrium is violated altogether and material from a possibly escaping exosphere may contribute to the absorption, as has been proposed for some lines in other planets \citep[][]{Zhang2022}.

All of these models considered here display similar absorption in both \ion{Na}{I}~D lines, which is hardly compatible with our observations. Moreover, for these models, the strength of the \ion{Na}{I}~D$_2$ line is underestimated, while that of the D$_1$ line is overestimated. Increasing the temperature or Na abundance
increases the absorption in both lines, but not sufficiently to reproduce the \ion{Na}{I}~D$_2$ signal
in the transmission spectrum.

\subsection{Comparison with known \ion{Na}{I} signals} 

In Table \ref{tab.parameters}, we compare key parameters of WASP-7 b to those of other systems with previously reported detections of absorption in the \ion{Na}{I}~D lines and reported Gaussian fit parameters. In particular, we give the equilibrium temperature of the respective planet, along with an estimate of the scale height, $H$, calculated assuming a mean molecular weight of $2.3$ for the atmosphere, and the resulting fractional area of the stellar disk, covered by the atmospheric annulus per scale height during transit:

\begin{equation}
    \Delta A = \frac{2\pi R_{\rm p} H}{\pi R_{\star}^2} \; ,
\end{equation}

where $R_{\rm p}$ and $R_{\star}$ are the planetary and stellar radii, respectively. Finally, we list the line contrast and FWHM of Gaussian fits to the observed transmission signal in the \ion{Na}{I}~D lines.

\begin{figure}[ht]
\centering
        \includegraphics[width = \columnwidth]{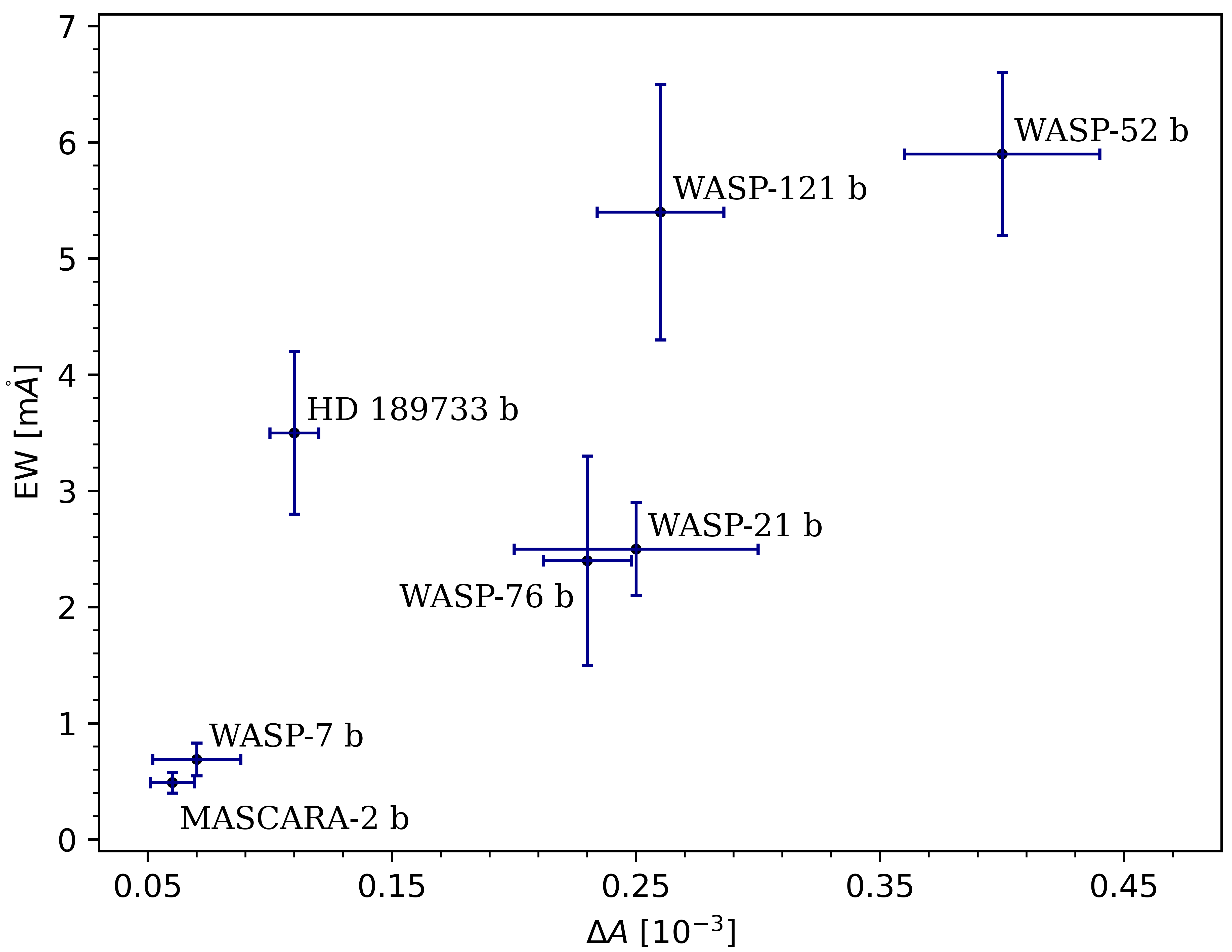}
        \caption{Excess absorption as indicated by Gaussian fits to the \ion{Na}{I}~D$_2$ line as a function of fractional atmospheric coverage $\Delta A$.}
        \label{fig:comparison}
\end{figure}

In Fig.~\ref{fig:comparison}, we compare the fractional atmospheric coverage, $\Delta A$, with the excess absorption as measured by the EWs of Gaussian fits to the D$_2$ line.
Among the listed exoplanets, WASP-52~b shows the largest fractional atmospheric coverage per scale height and the largest observed EW. On the other end of the scale, MASCARA-2 b exhibits the smallest fractional coverage, as well as a narrow absorption line with a low EW. For the considered systems, larger fractional coverage tends to be correlated with stronger absorption in the \ion{Na}{I}~D$_2$ line core, with a value of $0.82$ for Pearson's correlation coefficient. Although the low number of systems prevents a strong conclusion regarding the relation, the signal found in \wb\ fits well in the picture.

\begin{figure}
  \includegraphics[width = \columnwidth, height= 0.65 \columnwidth]{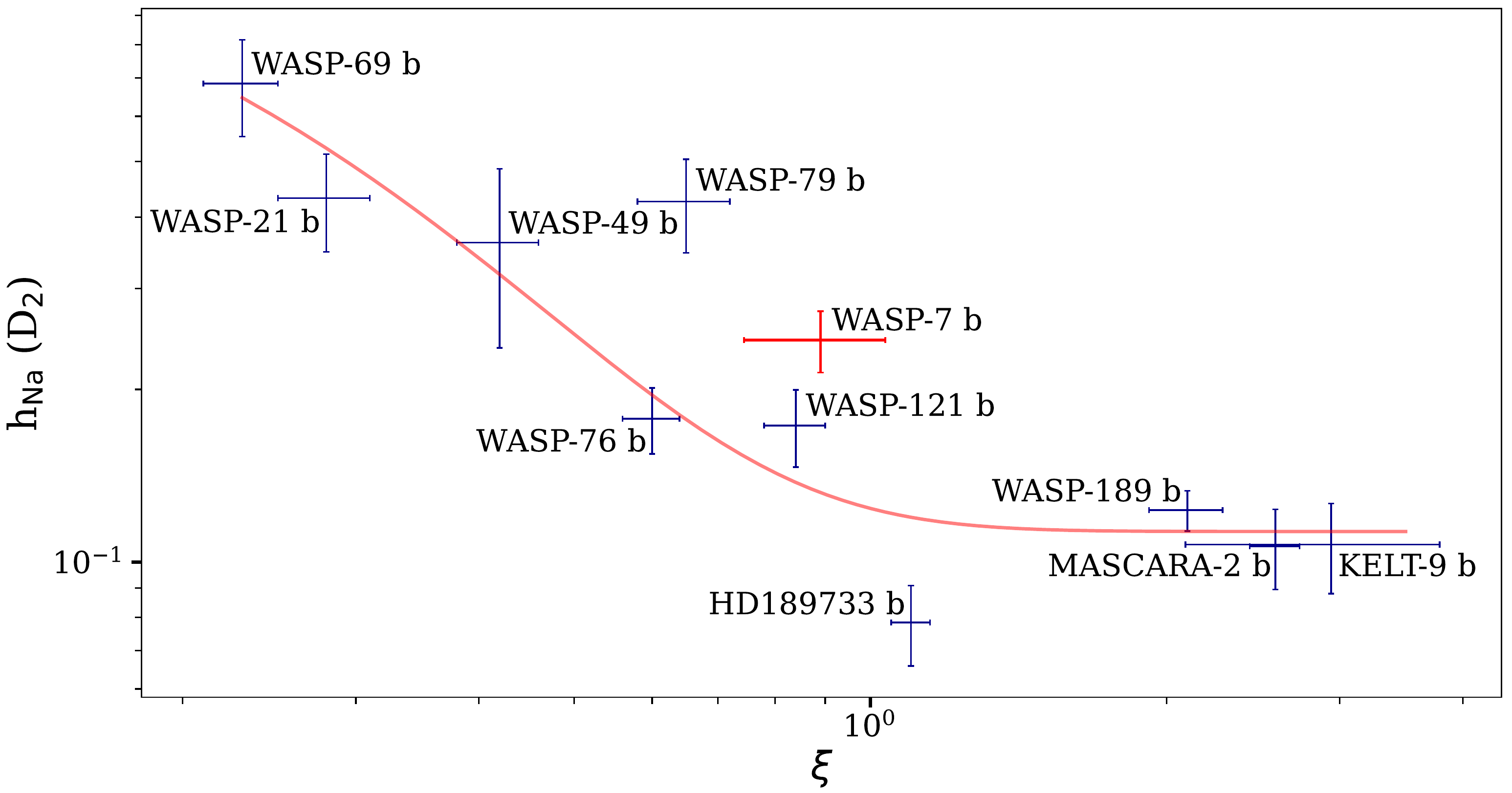}
    \caption{Height of atmosphere absorbing in the \ion{Na}{I}~D$_2$ line center divided by planetary
    radius as a function of $\xi$, along with the relation fitted by \citet{Langeveld2022}.}
    \label{fig.langf6}
\end{figure}

In their analysis of planetary sodium absorption in a sample of ten systems, which comprises the sample of Table~\ref{tab.parameters}, apart from WASP-7, \citet{Langeveld2022} point out an empirical relation between the height of the atmosphere absorbing
in the \ion{Na}{I}~D$_2$ line center divided by the planetary radius, $h_{\rm Na}$, and the quantity:

\begin{equation}
    \xi = \left(\frac{T_{\rm eq}}{1000~\mbox{K}} \right) \times \left(\frac{g}{g_{\rm J}}\right) \; ,
\end{equation}

where $T_{\rm eq}$ is the equilibrium temperature, $g$ is surface gravity, and $g_{\rm J}$ is the surface gravity of Jupiter. In Fig.~\ref{fig.langf6}, we use the data published by \citet{Langeveld2022}, which are weight combined, along with their best-fit relation, and add the data of WASP-7. The data points associated with WASP-7 lie above the best-fit relation given by \citet{Langeveld2022}, but fit reasonably well into the relation and are not outliers.

The FWHMs of the listed \ion{Na}{I}~D$_2$ line signals vary by a factor of about six, ranging from $0.13$~\AA\ to $0.73$~\AA, but, with the notable exception of the case of WASP-52~b, the widths of the \ion{Na}{I}~D$_1$ and D$_2$ lines listed in Table~\ref{tab.parameters} are consistent within two standard errors, which is in line with the modeling results we obtained for WASP-7\,b (Sect.~\ref{sec.modeling}). Of the cases shown in Table~\ref{tab.parameters}, WASP-7~b shows the lowest measurement of line width. The \ion{Na}{I}~D$_2$ transmission spectra of \wb\ and MASCARA-2\,b are quite similar in width and overall strength. However, in contrast to MASCARA-2\,b, where \citet[][]{CasasayasBarris2019} also detected a \ion{Na}{I}~D$_1$ line of comparable strength, we found a considerably weaker \ion{Na}{I}~D$_1$ line in \wb.

In their analysis, \citet{Langeveld2022} find that the ratio of contrasts of the \ion{Na}{I}~D$_2$ and D$_1$ transmission lines is consistent with one in seven systems and larger than one in three systems. Taking the values for WASP-7~b in Table~\ref{tab.parameters} at face value, we obtained ratios of $3.8\pm 1.3$ and $2.6\pm 0.5$ with the alternative analysis, discarding the ten post-transit spectra with the lowest S/N. For comparison, the largest ratio found by \citet{Langeveld2022} is $2.6 \pm 1.37$ (weight-combined spectra) for WASP-69~b, which is consistent with the numbers for WASP-7~b within the standard error. In principle, an overestimation of the \ion{Na}{I}~D$_2$ line contrast, an underestimation of the \ion{Na}{I}~D$_1$ line contrast, or both, could raise the ratio.

Excluding the ten post-transit spectra with the lowest S/N, which raises the measured \ion{Na}{I}~D$_1$ line contrast in our analysis, may indicate that the discrepancy may at least partially be caused by properties of the data possibly related to the S/N, or brought about by the inherent instability of the UVES instrument.

\begin{table*}
\centering
 \small
        \caption {Comparison of \wb\ parameters with those of other hot and ultra-hot Jupiters.}
        \begin{threeparttable}
                \begin{tabular}{l l l l l l l l l}
                        \hline\hline
                        \noalign{\smallskip}
                        Planet  & T$_{\rm eq}$ & H  \tablefootmark{a} & $\Delta A$ \tablefootmark{b} & D$_2$ line contrast & D$_2$ FWHM & D$_1$ line contrast & D$_1$ FWHM & Refs. \tablefootmark{d} \\
                                 & [K]  & [km] & [$10^{-3}$] & [\%] & [$\AA$] & [\%] & [$\AA$] \\
                        \hline
                        WASP-76 b &  2160  &  1479 & 0.23 & $0.37 \pm 0.09$ & $0.61 \pm 0.17$ & $0.50 \pm 0.08$ &  $0.68 \pm 0.12$ & 1, A \\
                        WASP-121~b &  2358 & 1011 & 0.26 & $0.69 \pm 0.12$ & $0.73 \pm 0.09$ & $0.25 \pm 0.09$ &  $0.9 \pm 0.1$ & 2, B \\
                        WASP-21 b & 1333 & 951 & 0.25 & $1.18$ $^{+ 0.23} _{-0.24}$ & $0.2$ $^{+ 0.04} _{-0.03}$  & $0.84$ $^{+ 0.16} _{-0.17}$ & $0.31 \pm 0.05$ & 3, C \\
                        WASP-52 b &  1315 & 672 & 0.4 & $1.31 \pm 0.13$ & $0.42 \pm 0.03$ & $1.09 \pm 0.16$ & $0.22 \pm 0.03$ & 4, D\\
                        \hline
                        WASP-7 b  &  1530 & 383 & 0.07 & $0.50\pm 0.06$ & $0.13\pm 0.02$ & $0.13\pm 0.04$ & $0.13\pm 0.02$ & 5, E \\
                        WASP-7 b (alt) \tablefootmark{c} &  & &  & $0.49\pm 0.05$ & $0.17\pm 0.02$ & $0.19\pm 0.03$ & $0.17\pm 0.02$ & 5, E \\
                        \hline
                        MASCARA-2~b &  2260  & 315 & 0.06 & $0.29 \pm 0.04$ & $0.16 \pm 0.02$ & $0.29 \pm 0.04$ & $0.15 \pm 0.02$ & 6, F \\
                        HD 189733 b &  1140 & 200 & 0.11 & $0.64 \pm 0.07$ & $0.52 \pm 0.08$ & $0.40 \pm 0.07$ & $0.52 \pm 0.08$ & 7, G \\
                        \hline
                \end{tabular}
                \label{tab.parameters}
                \tablefoot{
                \tablefoottext{a} {Atmospheric scale height.} \tablefoottext{b} {Fraction of the stellar disk covered by an atmospheric annulus with a height of one scale height.} \tablefoottext{c} {For WASP-7\,b we show the result of the modeling by considering all spectra and by ignoring ten post-transit spectra (alt).} \tablefoottext{d} {The numerals specify references for \ion{Na}{i}~D line contrasts and FWHMs, and letters denote references for the planetary effective temperature.} We calculated the scale heights based on the references.}
                \tablebib{
                (1)~\citet{Seidel2019}; 
                (A) \citet{West2016};
                (2) \citet{Cabot2020};
                (B) \citet{Delrez2016};
                (3) \citet{Chen2020a};
                (C) \citet{Ciceri2013};
                (4) \citet{Chen2020};
                (D) \citet{Hebrard2013};
                (5) {this work}; 
                (E) \citet{Southworth2012};
                (6) \citet{CasasayasBarris2019};
                (F) \citet{Talens2018};
                (7, G) \citet{Wyttenbach2015}.
 }
        \end{threeparttable}
\end{table*}

\section{Conclusions}
\label{sect:conclusion}

We used a high-resolution spectral transit time series obtained with UVES to study the transmission spectrum of the hot Jupiter WASP-7\,b, in particular the \ion{Na}{I}~D lines. We used telluric lines to account for instrumental drifts and carried out telluric correction using the \texttt{molecfit} package. To examine changes in the level of stellar magnetic activity, which might also affect the cores of the \ion{Na}{I}~D line, we investigated the cores of the \cahk,\ and hydrogen \ha\ lines, which exhibit marginal evolution. In particular, no flaring or distinguishable rotational modulation was detected.

We carried out transmission spectroscopy of the \ion{Na}{I}~D lines, and accounted for the CLV and RM effects by means of simulations. The resulting transmission spectrum shows a narrow absorption feature at the wavelength of the \ion{Na}{I}~D$_2$ line, for which we determined a line contrast of 0.50\,$\pm$\,0.06\,\% and a FWHM of 0.13\,$\pm$\,0.02\ {\AA,} close to the instrumental resolution with a FWHM of 0.09~{\AA}. Assuming the same width, we derived a line contrast of 0.13\,$\pm$\,0.04\,\% for the \ion{Na}{I}~D$_1$ line. Disregarding ten post-transit spectra with lower S/N (exposures 70-80), we obtained line contrasts of 0.19\,$\pm$\,0.03\,\% and 0.49\,$\pm$\,0.05\,\% for the \ion{Na}{I}~D$_1$ and D$_2$ lines, respectively, with a FWHM of 0.17\,$\pm$\,0.02\ {\AA}. Although consistent within the error, the so-determined line contrast of the \ion{Na}{I}~D$_1$ line is larger. We consider the detection for the \ion{Na}{I}~D$_1$ line absorption tentative, because the analysis of the BIC does not provide evidence for the presence of this line and, in our modeling, its width and shift are largely determined by the \ion{Na}{I}~D$_2$ line feature. Our investigation of the hydrogen Balmer lines (\ha, \hb, and \hg), the K~{\sc i} $\lambda$7699\,{\AA} line, the Ca~{\sc ii} H and K, and the IRT lines provided only upper limits for possible absorption features (Table~\ref{tab:upper-limit}).

Putting our results in the context of reports of \ion{Na}{I} absorption in other systems,  we find a good correlation between the fractional area of the stellar disk covered by an atmospheric annulus with a width of one scale height and the observed EWs of the \ion{Na}{I}~D$_2$ line features. Our results are consistent with this relation, as well as the empirical relation of \ion{Na}{I}~D$_2$ line strength described by \citet{Langeveld2022}. Among the considered systems, those with stronger absorption signals tend to indicate primarily broader absorption and only moderately deeper lines in the transmission spectrum.

For reference, we calculated a synthetic \ion{Na}{I} transmission spectrum using the pRT package, using an isothermal atmospheric model adapted to the WASP-7 system parameters. The model spectrum reproduced the observed line width well, but predicted overall weaker absorption. Both the pRT models, as well as observations in other systems such as MASCARA~2, also indicate significant absorption in the \ion{Na}{I}~D$_1$ line, compatible in strength with that in the D$_2$ line, which is hardly consistent with our results.
While we show that the absorption signal in the \ion{Na}{I}~D lines can plausibly be attributed to the planetary atmosphere, follow-up observations of WASP-7\,b are required to clarify the situation in the \ion{Na}{I}~D lines and improve the precision of the measurement to further constrain the structure of the atmosphere. 

\begin{acknowledgements}
        
We thank N. Casasayas-Barris and J.~Seidel for their useful discussion. We thank F. Pfeifer for helping to take the observations. SC acknowledges support by DFG through project CZ 222/5-1. SK acknowledges financial support from the DFG through priority program SPP 1992 “Exploring the Diversity of Extrasolar Planets” (KH 472/3-1).

\end{acknowledgements}

\bibliographystyle{aa} 
\bibliography{helib} 

\newpage
\appendix

\section{Best-fit parameters and transmission spectra}

We carried out the same analysis as mentioned in Sect.~\ref{sect:TSspectra}, and derived the result of the best-fit parameters and the transmission spectrum of \ion{Na}{I}~D lines by ignoring ten post-transit spectra (exposure 70-80). In addition, we obtained the transmission spectra of WASP-7 b for other lines, as described in Sect.~\ref{sect:upper-limit}.

\begin{table}[ht]
  \centering
        \caption {Best-fit parameters along with $68$\,\% credibility intervals of
        the transmission spectrum model 2.}
        \label{tab.fitTS}
    \begin{tabular}{l l l}
    \hline \hline
    Parameter & Value & Unit \\ \hline
    $P0$ & $0.0084\pm 0.0039$ \\
   $P1$ &  $0.0009\pm 0.0003$ & [$\AA^{-1}$] \\
    $s$ & $0.491\pm 0.050$ \\
    $RV$ & $ 0.836\pm 0.591$ &  [km$s^{-1}$] \\
    Contrast D$_2$ & $0.495 \pm 0.056$ &  [\%] \\
    Contrast D$_1$ & $0.191 \pm 0.039$ &  [\%] \\ 
    FWHM & $0.17 \pm 0.02$ & [$\AA$]  \\
    \hline
    Model & k & BIC \\ \hline
    No s & 6 & 1434.4 \\ 
    No Abs & 3 & 1536.1 \\ 
    Only~D$_2$ & 6 & 1378.1 \\ 
    Only~D$_1$ & 6 & 1533.3 \\
    D$_{1}$ and D$_{2}$ & 7 & 1359.0  \\
    \hline
    \end{tabular}
\end{table}

\begin{figure}[ht]
        \centering
  \includegraphics[width = \columnwidth, height= 0.6 \columnwidth]{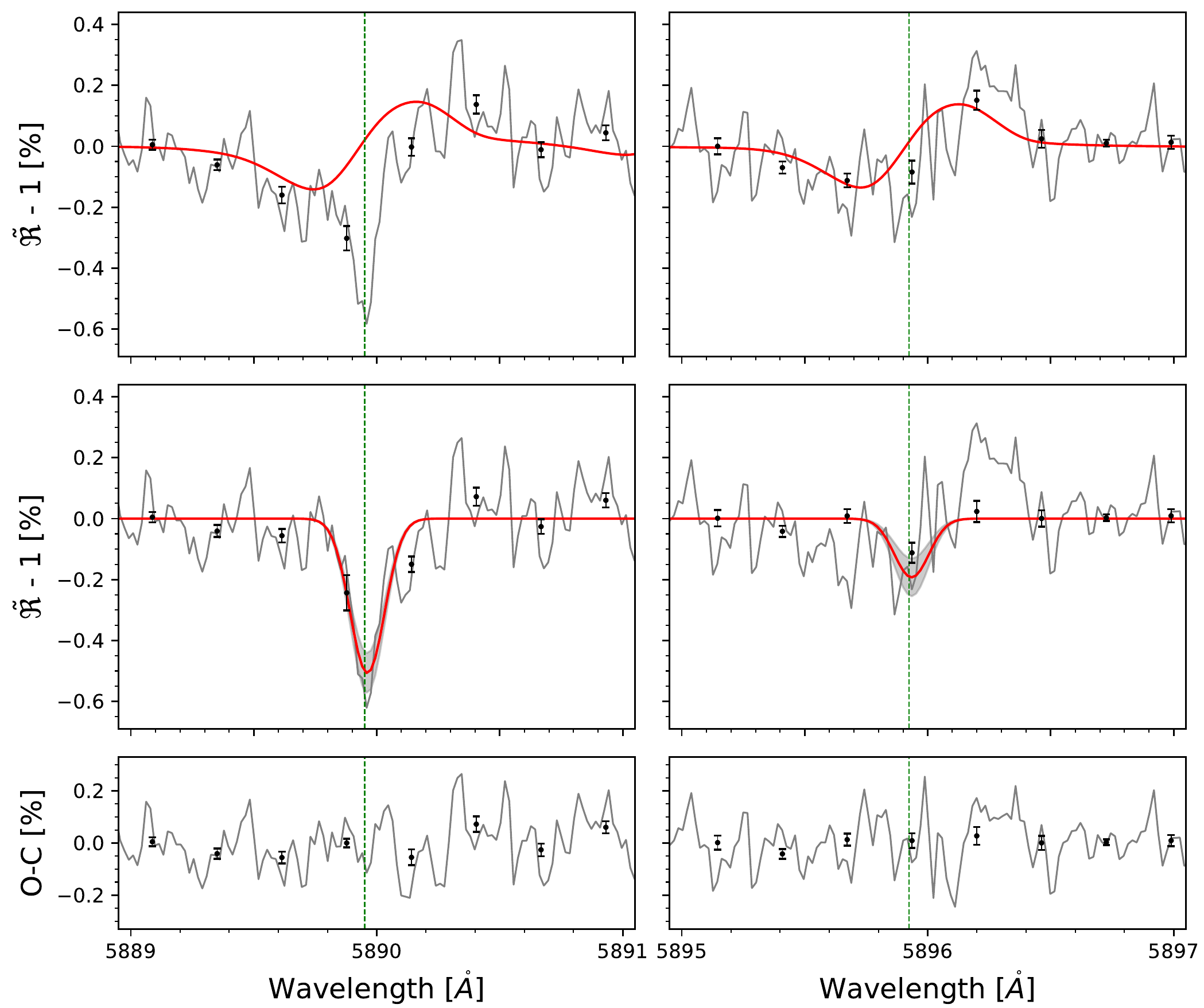}
        \caption{Transmission spectrum of WASP-7 b around the \ion{Na}{i}~D$_2$ (left column) and \ion{Na}{i}~D$_1$ line (right column) by disregarding ten post-transit spectra. Top panels: Observed transmission spectrum (gray) along with best-fit model components, representing the CLV and RM effects (solid red).
        Middle panels: Observed transmission spectrum with the best-fit Gaussian absorption components (red), and with the model shown in the upper panels subtracted (gray). The gray shades denote 1$\sigma$ uncertainty of the best-fit model (red). Bottom panel: Residuals with respect to the best-fit model. The black data points correspond to binning by a factor of 15. 
        }
        \label{fig:transmission-spectrum2}
\end{figure}

\begin{figure}[ht]
        \centering
  \includegraphics[width = \columnwidth, height= 0.6 \columnwidth]{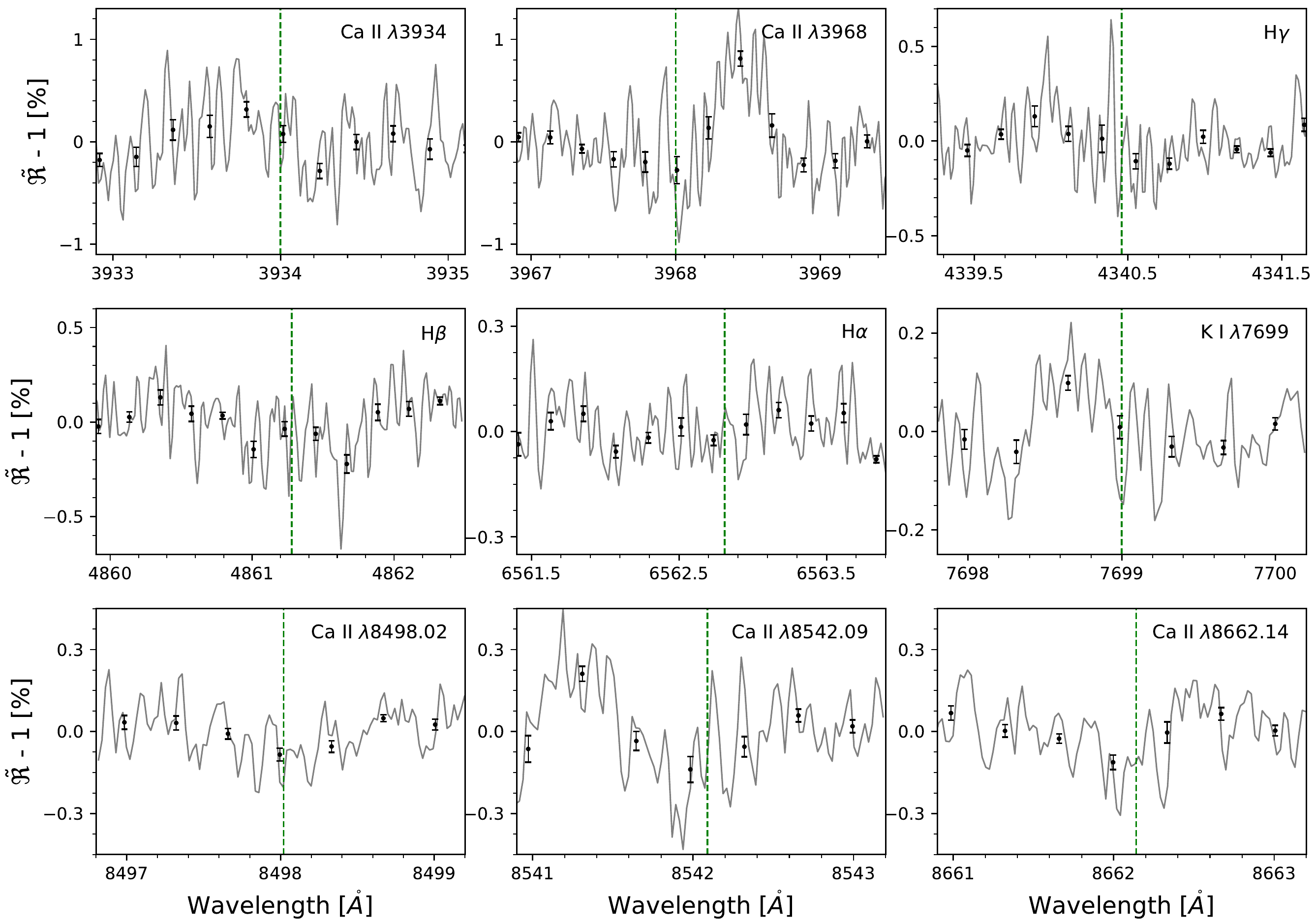}
        \caption{Observed transmission spectra of WASP-7 b for other lines (gray). Black data points correspond to binning by a factor of 15 and dashed vertical green lines indicate the centers of the lines.
        }
        \label{fig:TSs}
\end{figure}

\end{document}